\documentclass[twocolumn,superscriptaddress,prd,showkeys,showpacs,nofootinbib]{revtex4-1}

\usepackage{graphics}
\usepackage[colorlinks = true,
            linkcolor = cyan,
            urlcolor  = blue,
            citecolor = red,
            anchorcolor = blue]{hyperref}\usepackage{color}
\usepackage{amssymb}
\usepackage[nointegrals]{wasysym}
\usepackage{amsthm}
\usepackage{textcomp}
\usepackage{mathtools}
\usepackage{lineno}

\usepackage{comment}
\usepackage[separate-uncertainty,retain-explicit-plus,per-mode = symbol]{siunitx}
\usepackage{url}
\newcolumntype{C}[1]{>{\centering\arraybackslash}m{#1}}

\begin{document}

\title{Calibrating the scintillation and ionization responses of xenon recoils for high-energy dark matter searches}

\author{T.~Pershing} \email[Corresponding author, ] {pershing1@llnl.gov} \affiliation{Lawrence Livermore National Laboratory, 7000 East Ave., Livermore, CA 94551, USA}  
\author{D.~Naim} \affiliation{University of California Davis, Department of Physics, One Shields Ave., Davis, CA 95616, USA} 
\author{B.G.~Lenardo} \affiliation{Physics Department, Stanford University, Stanford, CA 94305, USA}  
\author{J.~Xu}\affiliation{Lawrence Livermore National Laboratory, 7000 East Ave., Livermore, CA 94551, USA}  
\author{J.~Kingston} \affiliation{University of California Davis, Department of Physics, One Shields Ave., Davis, CA 95616, USA}  
\author{E.~Mizrachi} \affiliation{Department of Physics, University of Maryland, College Park, MD 20742, USA}  
\author{V.~Mozin} \affiliation{Lawrence Livermore National Laboratory, 7000 East Ave., Livermore, CA 94551, USA}  
\author{P.~Kerr} \affiliation{Lawrence Livermore National Laboratory, 7000 East Ave., Livermore, CA 94551, USA}  
\author{S.~Pereverzev} \affiliation{Lawrence Livermore National Laboratory, 7000 East Ave., Livermore, CA 94551, USA}  
\author{A.~Bernstein} \affiliation{Lawrence Livermore National Laboratory, 7000 East Ave., Livermore, CA 94551, USA}  
\author{M.~Tripathi} \affiliation{University of California Davis, Department of Physics, One Shields Ave., Davis, CA 95616, USA} 


\begin{abstract}
Liquid xenon-based direct detection dark matter experiments have recently expanded their searches to include high-energy nuclear recoil events as motivated by effective field theory dark matter and inelastic dark matter interaction models, but few xenon recoil calibrations above 100 keV are currently available. 
In this work, we measured the scintillation and ionization yields of xenon recoils  up to 426 keV. The experiment uses 14.1 MeV neutrons to 
scatter off xenon in a compact liquid xenon time projection chamber
and produce quasi-monoenergetic xenon recoils between 39 keV and 426 keV.  
We report the xenon recoil responses and their electric field-dependence for recoil energies up to 306 keV; due to the low event statistics and the relatively mild field dependence, the yield values at higher energies are reported as the average of xenon responses for electric fields between 0.2-2.0 kV/cm. 
This result will enable xenon-based dark matter experiments to significantly increase their high energy dark matter sensitivities by including energy regions that were previously inaccessible due to lack of calibrations.  
\end{abstract}


\maketitle



\section{Introduction}
\label{sec:introduction}
Dual phase xenon Time Projection Chamber (TPC) experiments have become one of the leading technologies in searches for particle dark matter interactions, especially those expected of Weakly Interacting Massive Particles (WIMPs)~\cite{LUX2016_Run3_4,PandaX2017_DM,PandaX-4T:2021bab,ZEPLIN_Lebedenko:2008gb,XENON1T_2018, Xenon10_collaboration_search_2011,XENON2012_225dResult,LUX-ZEPLIN:2022qhg}. 
A typical xenon TPC focuses on the detection of nuclear recoils in the energy region of a few keV to tens of keV, where most signals from standard WIMP interactions are expected \cite{Jungman:1995df,Schumann:2019eaa}.
However, despite rapid improvement of experimental sensitivities in this energy region over the past few decades, no definitive evidence of dark matter interactions has been observed, leading to diminishing parameter space allowed for the standard WIMP model \cite{Escudero:2016gzx}.

Recent years have seen a growing interest in expanding the dark matter search energy range to include higher energy events, up to hundreds of keV energy for nuclear recoils, mainly motivated by the Effective Field Theory (EFT) dark matter interaction~\cite{Fan:2010gt,Fitzpatrick:2012ix} and inelastic dark matter interaction models~\cite{Smith_inelastic_2001}.
EFT is a generalized approach to treat WIMP-nucleon interactions beyond the simple spin-dependent and spin-independent couplings, and thus allows for more possible dark matter interaction modes to be experimentally tested. 
In this framework, all WIMP interactions with protons and neutrons that conform to  conservation laws are considered, including those involving the momentum and velocities of the interaction particles. 
As demonstrated in \cite{LUX:2020oan}, some of these new EFT operators can lead to the relative suppression of low-energy nuclear recoil signals following WIMP scatters, which can cause such dark matter interactions to evade detection in experiments that focus on the low-energy regions only.
Meanwhile, these WIMP interactions can still produce detectable nuclear recoil signals in the hundreds of keV energy region, which can be pursued and tested in existing detectors. 
The inelastic dark matter interaction model hypothesizes two nearly degenerate energy levels for the dark matter particles and can demonstrate a similar suppression of low-energy nuclear recoil interaction rates compared to those at higher energies \cite{Smith_inelastic_2001}. 

A few experiments have explored these dark matter interaction models and excluded new parameter space beyond what had been previously studied \cite{LUX:2021ksq,XENON:2019gfn,PandaX2018_EFT,DS50_2018_S2Only}.  Although dark matter experiments typically target interaction energies below a few 10's of keV, liquid xenon TPCs can retain high signal acceptance for interactions in higher energy regions and can readily expand their dark matter search scopes. 

Still, expanding dark matter searches to high energy windows faces technical challenges, with the most important one being the uncalibrated detector responses in the new energy regime.
A xenon TPC collects both scintillation photons and ionization electrons, which must be converted to xenon recoil energy before the observed events can be tested against dark matter-induced signal models. 
The response of liquid xenon to nuclear recoils has been extensively studied in the energy region up to around 100 keV using neutrons to mimic dark matter interactions \cite{Lenardo2015_XeLightCharge}. 
For energies above 100 keV, however, few calibrations exist, and existing measurements have not adequately studied the effect of the electric field strength at the interaction site. 
Since different experiments often operate at different electric fields, 
it is necessary to systematically study the field effects to accurately model the detector response to high energy dark matter interactions. 

In this article, we report a calibration of liquid xenon response to nuclear recoils between 39 keV and 426 keV. 
To our knowledge, this is the highest energy xenon recoil calibration result reported to date. 
In Sec.~\ref{sec:setup} we describe the experimental setup and the measurement technique, and in Sec.~\ref{sec:analysis} we discuss the data analysis methods. 
The scintillation and ionization yields of xenon recoils extracted from the measurement are explained in Sec.~\ref{sec:discussion}, and we conclude in Sec.~\ref{sec:conclusion}.

\section{Experimental setup}
\label{sec:setup}
A compact xenon Time Projection Chamber (TPC) is used for this experiment.
The active liquid xenon volume ($\diameter$ 38 mm $\times$ 25 mm) is enclosed in a 2.5 mm thick, high-reflectivity PTFE cylinder to enhance the  collection of liquid xenon scintillation light (referred to as S1). The PTFE cylinder consists of three segments separated by two electric grids (cathode and extraction grid) and is nested inside a field-cage that provides the required electric field for the TPC operation. 
The TPC anode grid (grounded) is installed 5.75 mm above the liquid surface to ensure a uniform electric field at the liquid surface and in the gas. 
Ionization electrons produced by particle interactions between the cathode and extraction grid are drifted upwards to the liquid surface and subsequently extracted into the gas phase, where they produce secondary electroluminescence photons (referred to as S2).  
Four 1" (2.54 cm) Hamamatsu R8520-406 PMTs are installed right above the anode grid and a single 2" (5.08 cm) Hamamatsu R8778 PMT is installed $\sim$10 mm below the cathode to collect the S1 and S2 light signals. 
An additional grounded grid right above the bottom PMT shields the PMT from the TPC high voltage in a similar fashion that the anode shields the top PMT array. 
A detailed description of the apparatus can be found in \cite{Lenardo:2019fcn}.

In the first part of the experiment, a high voltage of -12 kV was applied on the TPC extraction grid, leading to an extraction field of $7.2\pm0.1$ kV/cm (calculated using COMSOL) and an average electron extraction efficiency of 97.5\% \cite{Xu:2019dqb}. 
The voltage was later lowered to -11 kV (extraction field of $6.6\pm0.1$ kV/cm, extraction efficiency 95.6\%) to reduce the saturation risk of high energy S2 signals.
The TPC cathode voltage was set at 0.5kV, 1.9kV and 5kV below the extraction grid voltage in different data sets so the drift field dependence of xenon recoil response can be evaluated.  
These cathode voltages corresponded to drift fields of $200\pm45$ V/cm, $760\pm56$ V/cm, and $2000\pm81$ V/cm in the active xenon volume. 

Nuclear recoil events in the xenon TPC are produced by bombardments of 14.1 MeV neutrons from a deuterium-tritium (DT) neutron generator (Thermo Scientific P385). 
During most of the experiment, the DT generator was operated at an output of 4.7e7 neutrons per second.
To produce a collimated neutron beam and to reduce the radiation hazard from the high-energy neutrons, a customized neutron shielding structure was constructed. 
First, 8 inches of lead surrounds the neutron emission spot to down-shift the off-beam neutron energy via the (n,2n) interaction. 
A 2.54 cm diameter hole in the lead shielding produces a narrow neutron beam perpendicular to the DT generator and toward the xenon TPC. 
The DT source and lead shielding are then surrounded by a 
cube of borated water shielding (saturated Borax water solution, 0.36\% B) constructed with 
interlocking water containers to moderate and absorb the off-beam neutrons. 
Between the DT generator and the TPC, a collimation hole was formed inside the water structure using 
six sheets of 30\% borated polyethelene with a 2.54 cm $\times$ 2.54 cm square hole aligned with the lead shielding collimator to allow for neutron passage.

\begin{figure}[!h]
    \centering
    \includegraphics[width=0.49\textwidth]{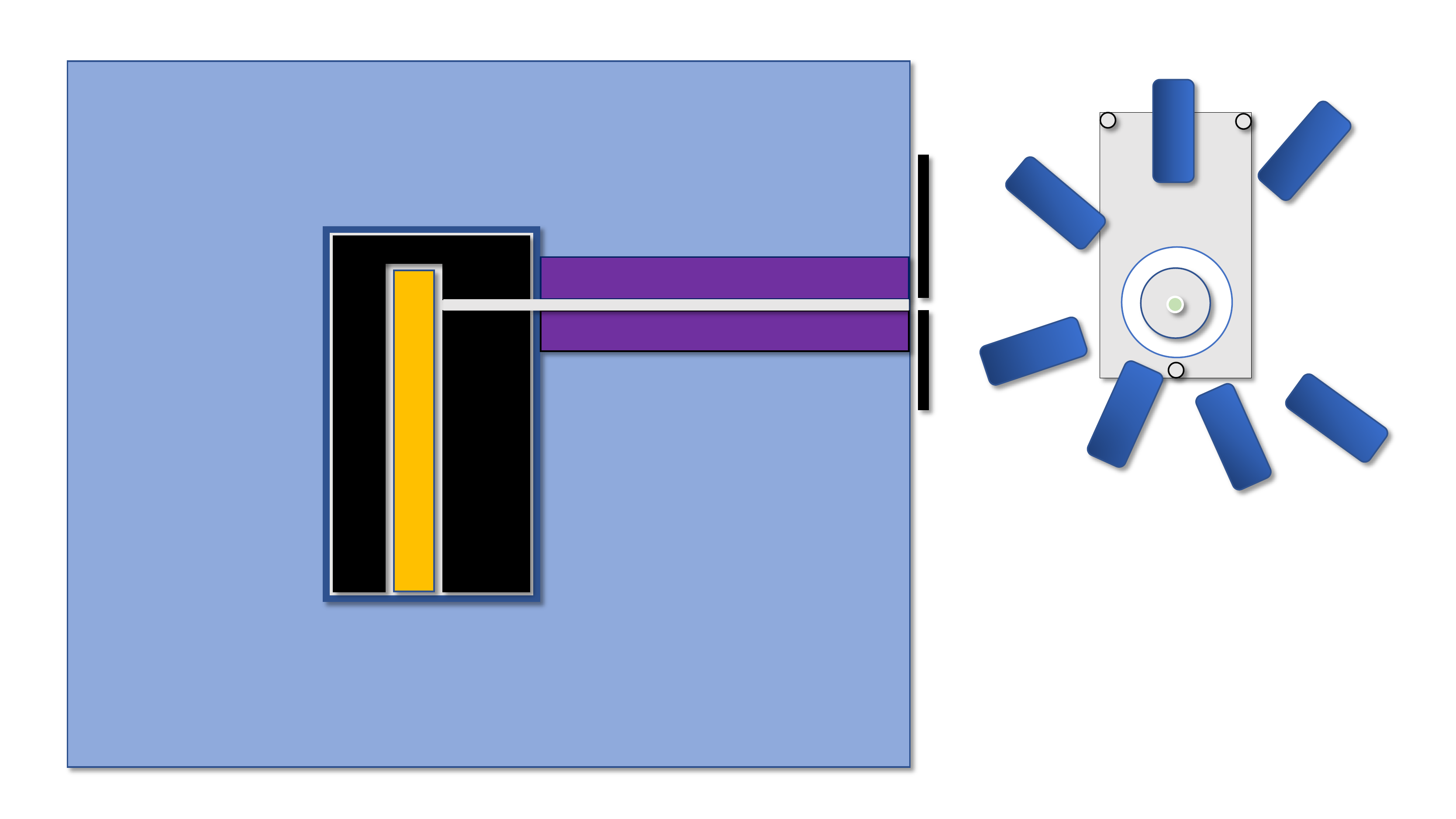}
    \caption{Layout of the high-energy xenon recoil experimental setup.  The DT source (orange) is surrounded with a layer of lead (black) and a borated water shield (240 cm $\times$ 190 cm $\times$ 150 cm, light blue).  A collimator (white) formed in the shielding using borated polyethylene (purple) leads directly from the DT neutron source to the xenon TPC (described in text).  Seven LS backing detectors (dark blue) are placed at various locations surrounding the xenon TPC to detect scattered neutrons. }
    \label{fig:setup}
\end{figure}

As illustrated in Fig.~\ref{fig:setup}, 7 Liquid Scintillator (LS) backing detectors are positioned around the xenon TPC to tag neutrons with scattering angles ranging from 36 to 162 degrees with respect to the collimated neutron beam.  
For neutrons that undergo single scatters in the xenon detector, the energy transfer from neutrons to xenon nuclei is kinematically constrained by the incoming neutron energy and the scatter angle. Therefore, xenon recoils in the TPC that are in coincidence with a specific LS detector will form a quasi-monoenergetic distribution, with the energy spread rising from the finite size of the xenon active volume and the LS detector volume. 
The LS detectors in this experiment are placed at such angles that xenon recoils between 39 and 426 keV energy can be tagged. 
Because the elastic scatter cross section of 14.1 MeV neutrons with xenon decreases at high scatter angles, we placed the large-angle LS detectors closer to the xenon TPC to enhance the event rates, at the cost of slightly increased angular spread.   
The distance between LS detectors and the xenon TPC ranged from 35.4 cm at the 36 degree scattering angle to 17.3 cm at the 115 degree scattering angle. All LS detectors have a cylindrical target volume of 10.2 cm diameter and 7.6 cm depth coupled to a single PMT, and are filled with EJ-301 or EJ-309 scintillator from Eljen.

The TPC and LS data acquisition system is triggered by a 30 $\mu$s-coincidence  between the LS trigger, defined as the OR logic among all LS detector signals, and the TPC trigger, defined as a three-fold coincidence between the four top PMT signals.
To reduce the random coincidence trigger rate, each LS trigger threshold was set to be  near 10\% of the maximum neutron energy deposition in this backing detector, and the TPC trigger threshold was set to be slightly higher than the amplitude of a single ionization electron.
For each trigger, we record 60 $\mu$s of continuous waveform for each TPC PMT and LS detector output, starting from 30 $\mu$s before the trigger time.  Signals were digitized at 250 MHz by a 14-bit Struck SIS3316 digitizer. 
A total of $\sim$50 hours of data was acquired in this experiment. 

\section{Data Analysis}
\label{sec:analysis}
The recorded event waveforms are analyzed offline using a customized data analysis package that was adapted from DAQMAN \cite{DAQMAN}.
First the baseline is estimated for each raw PMT waveform and subtracted, allowing the PMT signals to be extracted.  
For the TPC PMT signals, the single photoelectron (SPE) peaks in the tail of large pulses~\footnote{Events with pulses that saturate the PMTs or the electronics are excluded from the SPE calibration.} are used to estimate the PMT gain as an absolute calibration.  For the LS detectors, relatively low PMT gain values were used to avoid saturation of high-energy neutron signals; as a result, no SPE calibration could be obtained and we used a relative energy calibration that placed the maximum neutron interaction energy in different detectors at approximately the same numerical value. 
Afterwards, pulses are identified in the summed TPC channel, which is the summation of all SPE-calibrated TPC waveforms, and in the individual LS detector channels. 
For each identified pulse, information including the pulse time, area, amplitude, and pulse shape characterization parameters are saved for analysis. 
Hundreds of events were manually scanned to ensure proper pulse finding, and no significant efficiency loss from pulse finding was observed for the lowest energy events considered in this analysis. 

\subsection{Neutron scatter event selection}\label{subsec:EventSelection}

For a recorded TPC-LS detector coincidence event to be considered a valid neutron scatter candidate, it has to satisfy the following event selection criteria: 
\begin{itemize}
    \item The TPC signal needs to be a single scatter nuclear recoil within the active volume. 
    \item  Only one LS detector should observe a signal in the coincidence time window, and the signal's total area and pulse shape discrimination (PSD) parameter must be consistent with a neutron interaction.
    \item The time difference between the TPC signal and the LS signal should be consistent with the neutron time of flight (TOF) between the two detectors. 
\end{itemize}

Each candidate xenon TPC event is required to have a valid S1 and S2 pair.  S1 and S2 pulses are first identified using TPC pulse width and top-bottom asymmetry classifiers (see \cite{Xu:2019dqb} for more details).  S1 scintillation signals (width of $\sim$10s of nanoseconds from scintillation de-excitation time) are typically much narrower than S2 ionization signals ($\sim$microsecond width due to electron transport in the gas phase of the TPC), allowing pulse width characterization using the time difference between 10\% of the pulse integral and 25\% of the pulse integral.  Additionally, for the TPC used in this work, S1s originating from the drift region deposit $\sim$85\% of their light in the bottom PMT due to total internal reflection on the xenon liquid surface, while S2s originating from the gas phase deposit approximately half of their light in the top and bottom PMT arrays.  When more than one S2 pulse is identified in an event, we require the first S2 to also be the largest S2 in the event, and that the total pulse area in the S1 and the first S2 should contain $>$80\% of the total TPC waveform area.  
This cut removes most neutron multiple-scatter backgrounds while preserving valid events in which secondary S2s are produced by the primary S2 through photoionization \cite{LUX:Ebkg2020vbj,XENON:Ebkg2021myl}.

\begin{figure}[!ht]
    \centering
    \includegraphics[width=0.50\textwidth]{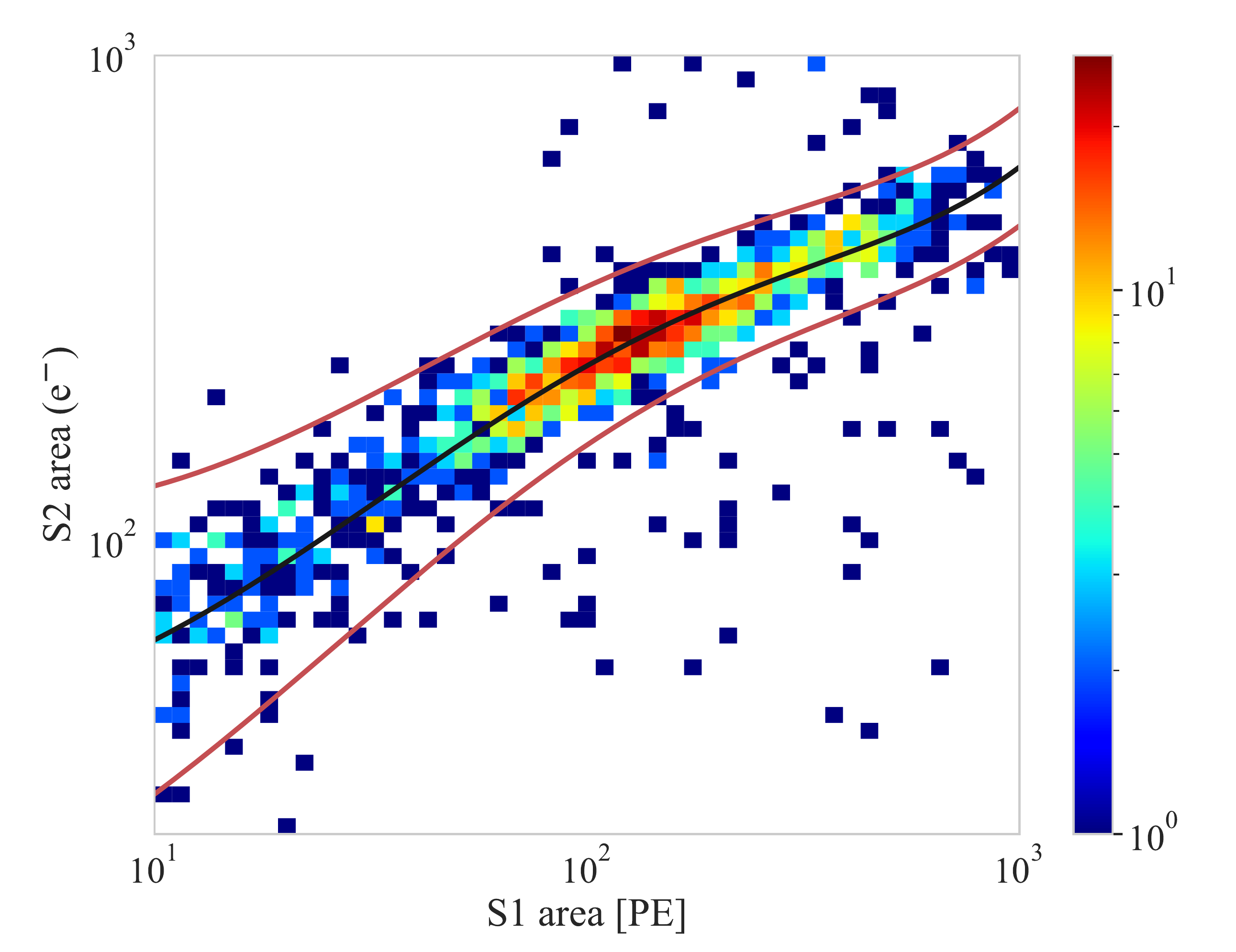}
    \caption{S2 vs. S1 pulse area distribution for all DT neutron scatter candidates at all recoil energies for data with a TPC drift field of 2000 V/cm and liquid extraction field of 6.6 kV/cm.  All preliminary cuts have been applied except the NR band cut.  The black and red lines show the mean and $\pm3\sigma$ range of the nuclear recoil band measured in TPC trigger-only data, which is used as the nuclear recoil selection band at this field configuration.} 
    \label{fig:s12}
\end{figure}

The identified S1-S2 pair can then be used to estimate the position of the interaction. Only events with a S1-S2 time separation (aka, electron cloud drift time) that corresponds to interaction positions between the TPC cathode and extraction grid are accepted. 
The minimum allowed drift time was chosen to be 3.5 $\mu$s to exclude interactions occurring near or above the extraction grid.  The maximum drift time cut was set to exclude events within a few millimeters of the cathode, with maximum drift time cuts of 16.2, 13.7, and 12.0 $\mu$s used for data with 200, 760, and 2000 V/cm drift voltage settings, respectively. 
No radial position cut was applied in this analysis, as both the S2 and S1 area for nuclear recoil events were found to be uniform as a function of reconstructed XY position.  This uniformity is a result of fiducialization already provided by the PTFE reflector inside the TPC. 

Nuclear recoil and electron recoil interactions in liquid xenon lead to different energy partitioning between scintillation and ionization energy channels  
\cite{Aprile2007_XeScintIon}. 
We select xenon nuclear recoil events in the active volume with a band cut in the S1-S2 parameter space, as illustrated in Fig.~\ref{fig:s12}. 
Empirical cut values are derived for each TPC drift field and extraction field configuration using high-statistics nuclear recoil data triggered by the TPC only (no coincidence with the LS detectors were required), which conservatively retains $>$98\% of all nuclear recoil events in the TPC-BD coincidence data used for the primary analysis.
Although this empirical nuclear recoil band definition could be subject to biases from features in the recoil energy spectrum, this effect may only modify the signal acceptance at the percent level and is not expected to impact the evaluation of mean scintillation and ionization yield values in this work. 

\begin{figure}[!ht]
\centering
\begin{minipage}{.50\textwidth}
  \centering
  \includegraphics[width=0.95\linewidth]{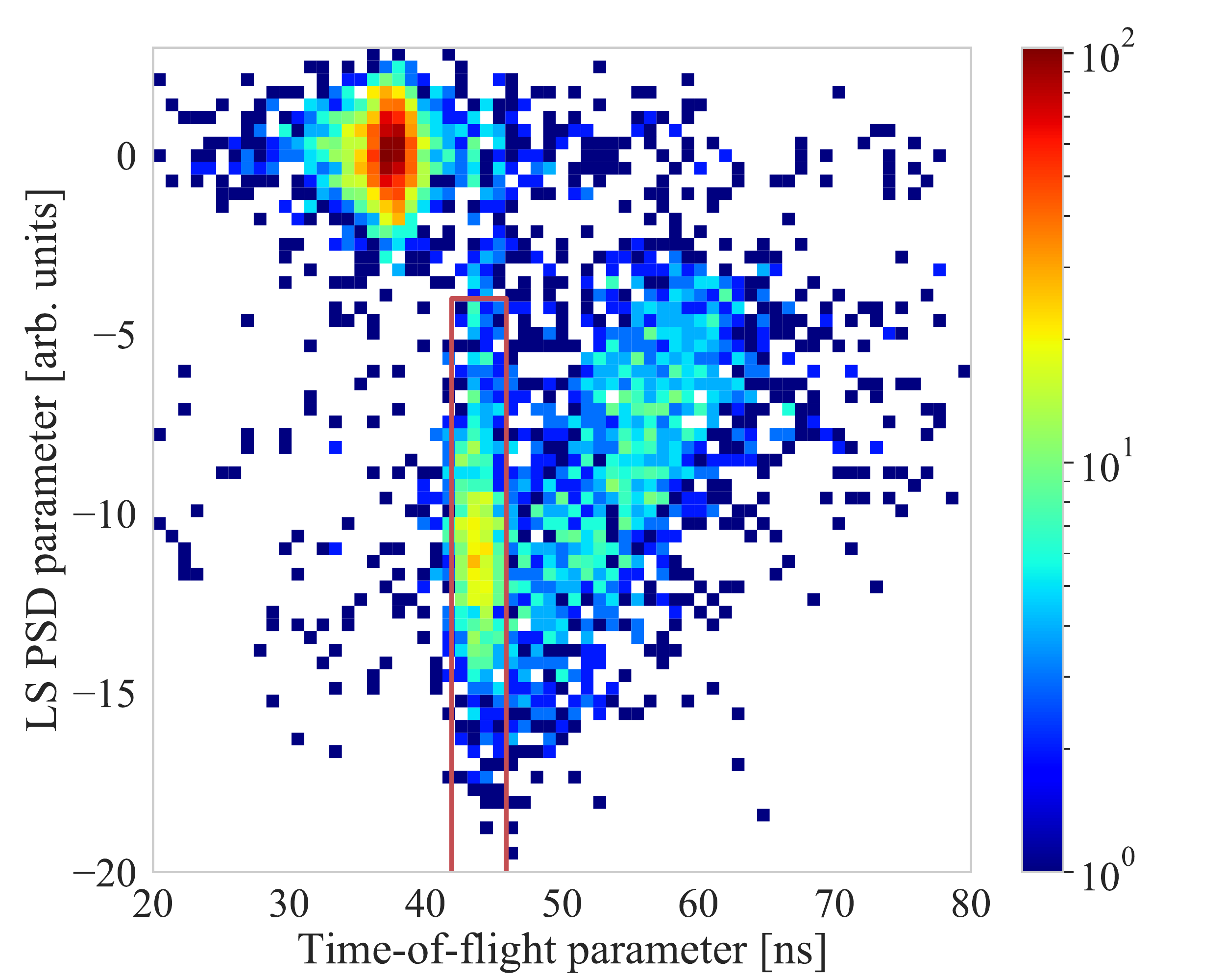}
\end{minipage}
\begin{minipage}{.50\textwidth}
  \centering
  \includegraphics[width=0.95\linewidth]{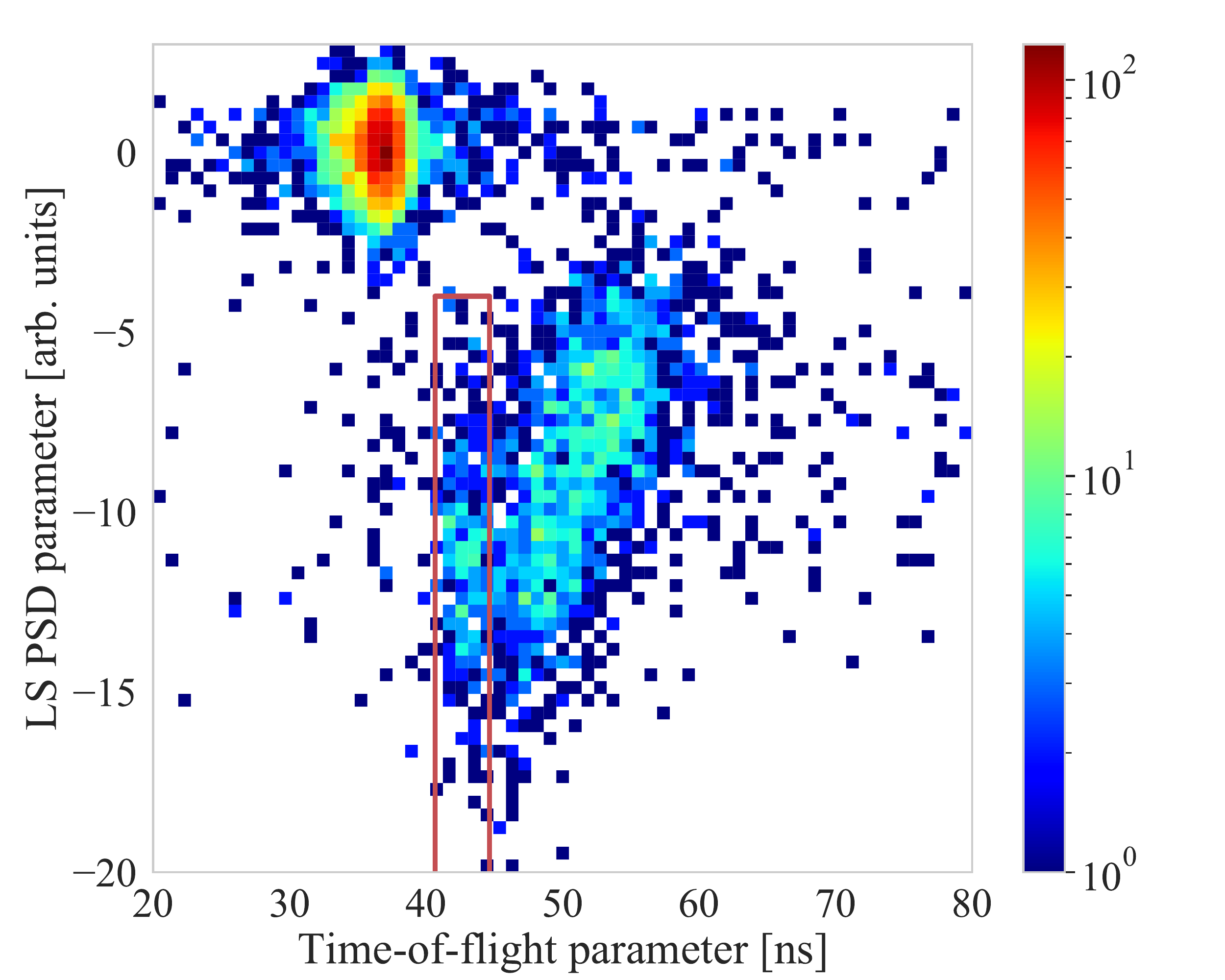}
\end{minipage}%
\caption{Backing detector PSD vs. Time-of-flight parameter for recoil events in LS backing detectors corresponding to 75 keV (top) and 219 keV (bottom) xenon recoil energies.  The red lines show the TOF window and LS PSD parameter cuts used to select primary nuclear recoil events, which form a peak in the boxed regions.  Gamma events form the top-left populations with earlier arrival time and larger LS PSD parameter (explained in text), while lower-energy secondary scatter neutrons form the continuous distributions at larger time-of-flight.  These key features are broadly present and similar at all recoil energies, with the primary nuclear recoil peak reducing in intensity as the recoil energy increases and cross-section decreases.}
\label{fig:tofpsd}
\end{figure}

Once a valid TPC signal is identified, we require that one and only one prominent LS detector pulse is in coincidence with the TPC signal. 
When more than one LS pulse is found in the coincidence window, the largest pulse must account for $>$90\% of the total energy recorded in all LS detectors for the coincidence event to be considered usable. For the prominent LS pulse, two cuts were then applied to select neutron-like signals.  First, the LS pulse area needs to reside within 10-110\% of the LS detector's neutron pulse area distribution endpoint, which was characterized separately for each LS detector.
Then, a pulse shape discrimination (PSD) parameter was used to classify LS signals as gamma-like or neutron-like.  A PSD parameter measuring the prompt-to-total charge ratio is used to characterize LS pulses, where the prompt window is defined as the first 50 nanoseconds of the pulse.  
As shown in Fig.~\ref{fig:tofpsd}, the LS PSD parameter is normalized as the distance from the gamma PSD band mean in the unit of the standard deviation of the gamma band for the corresponding recoil energy. As a result, gammas always have a mean PSD value of zero, and neutron events always have negative PSD values.  For all recoil energies, an LS signal is accepted as a neutron-like signal for PSD values $<$-4.  This cut ensures that few prompt gamma events contaminate the NR candidate event dataset while conservatively accepting $>$90\% of neutron recoils in the LS detector for the selected energy range.

The final selection cut takes advantage of the different traveling speeds for neutrons and gammas. 
At a kinetic energy of 14.1 MeV, neutrons travel at a speed of approximately 5 cm/ns, which is several times lower than that of gammas (around 30 cm/ns). 
Therefore, we can use the time separation between the nuclear recoil interaction in the TPC (measured by the S1 time) and the neutron detection time in the LS detector, defined as neutron time-of-flight (TOF), to select true neutron coincidence events. 
Fig.~\ref{fig:tofpsd} shows two example distributions of LS PSD and TOF parameters for both neutron and gamma coincidence events, where the gammas exhibit higher PSD ($\sim$0) and lower TOF while the neutron population has a lower PSD value ($<$-4) and higher TOF.  The TOF separation between the gamma population and the neutron population is 3.5-7 ns for different LS detectors, consistent with the inter-detector distances of 17.3--35.4cm. 
A TOF spread of $\sim$1ns was achieved in this experiment through a linear interpolation between subsequent digitizer samples. 

A population of neutron events with larger TOF values than that expected of 14.1 MeV neutrons~\footnote{The maximum neutron energy transfer to xenon in a single scatter is $\sim$420keV, which has a negligible effect on the neutron TOF distribution. }  is also observed in Fig.~\ref{fig:tofpsd}.  
As the TOF value increases within this population, the PSD value also becomes more and more gamma-like, consistent with that expected for reduced energy neutrons.
These events are explained as spallation and inelastic scattering of the 14.1 MeV neutrons in the xenon target~\footnote{It is worth noting that these interactions also produce gammas, which contributed to a large fraction of the coincident gammas detected in the LS detectors.}, for which the total cross section is comparable to that of elastic scattering.  
Due to the low elastic scattering cross section of 14.1 MeV neutrons at angles greater than 90 degrees, the low energy neutron background begins to overshadow the single scatter neutron population in the TOF distribution, as illustrated in Fig.~\ref{fig:tofpsd}, bottom.
In these cases, we define the TOF cut based on the calculated TOF values, which was verified to be accurate at the 0.1 ns level based on comparison with the time-of-flights directly measured for low-angle scatter events. 
The TOF window used in this work is $\pm2\sigma$ from the central neutron TOF value. 
As discussed in Sec.~\ref{subsec:systematics}, the systematic uncertainty due to contamination of slow neutron interactions was evaluated by varying the TOF cut values.

\subsection{TPC signal collection efficiency}
\label{subsec:EEELCE}

In this measurement, both the scintillation and ionization signals produced by xenon recoils in liquid xenon are collected as PMT pulses. To convert the measured S1 and S2 pulse area to the number of primary photons and electrons released at the interaction site, we need to independently evaluate the collection efficiencies of scintillation light and ionization electrons in the TPC.

\begin{figure}[!ht]
    \centering
    \includegraphics[width=0.43\textwidth]{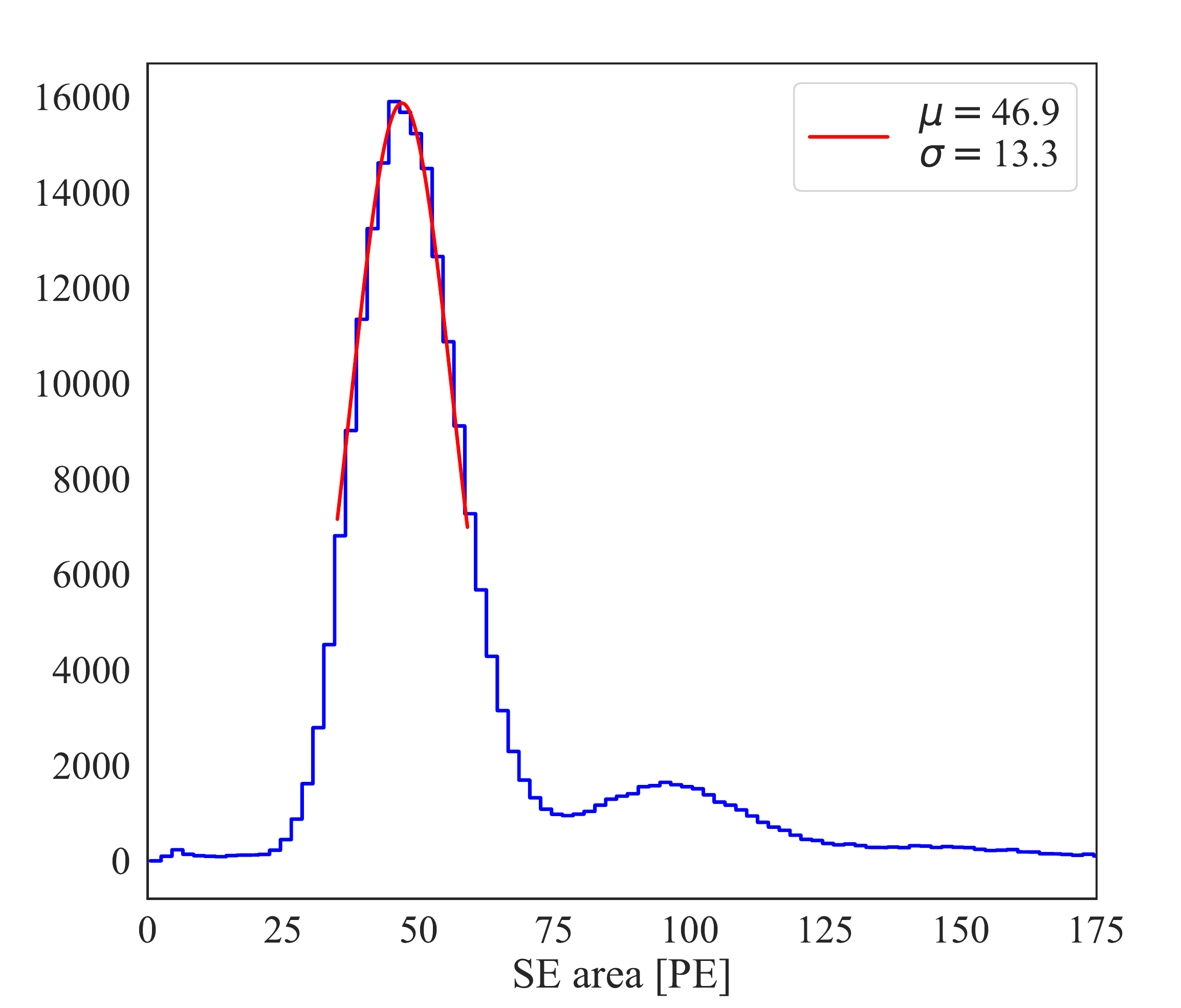}
    \caption{Single electron area distribution in units of photoelectrons.  The single electron area is measured using the top PMT array alone.  The single electron size shown was measured during operation with the extraction grid voltage at -12 kV.}
    \label{fig:se}
\end{figure}

For the ionization signals, the TPC has sufficient amplification gain to enable the definitive detection of single ionization electrons for the neutron scatter events. 
Fig.~\ref{fig:se} shows an example distribution of single electron (SE) pulse area recorded by the top PMT array at 7.2 kV/cm liquid extraction field, which can be well approximated as a Gaussian with a mean of 46.9 p.e. and a width of 13.3 p.e.
The bottom PMT observes a similar number of photoelectrons for S2 pulses as the sum of the top four PMTs, but the large pulse area concentrated in a single PMT channel tends to saturate the readout electronics especially for high-energy S2s.
In addition, the bottom PMT gain calibration exhibits a $\sim$10\% uncertainty when measured using photons following S2s of different areas.  For these reasons, we exclude the bottom PMT signals from the S2 size evaluation.

The detected number of ionization electrons calculated from the S2 signal size and SE calibration suffers from the electron loss to captures by impurities in the liquid as well as the efficiency to extract an electron from the liquid into the gas.
Thanks to the small size of the TPC and the continuous purification of the xenon through a hot getter during detector operation, we find that the S2 signal size is constant within uncertainties across the TPC drift region.  This suggests 
the electron loss to impurities is at the percent-level or smaller, which is subdominant to our statistical and systematic uncertainties, and can be neglected in this work.
The electron extraction efficiency is corrected for using values discussed in Sec.~\ref{sec:setup}.

\begin{figure}[!ht]
\centering
\begin{minipage}{.45\textwidth} 
  \centering
  \includegraphics[width=1.02\linewidth]{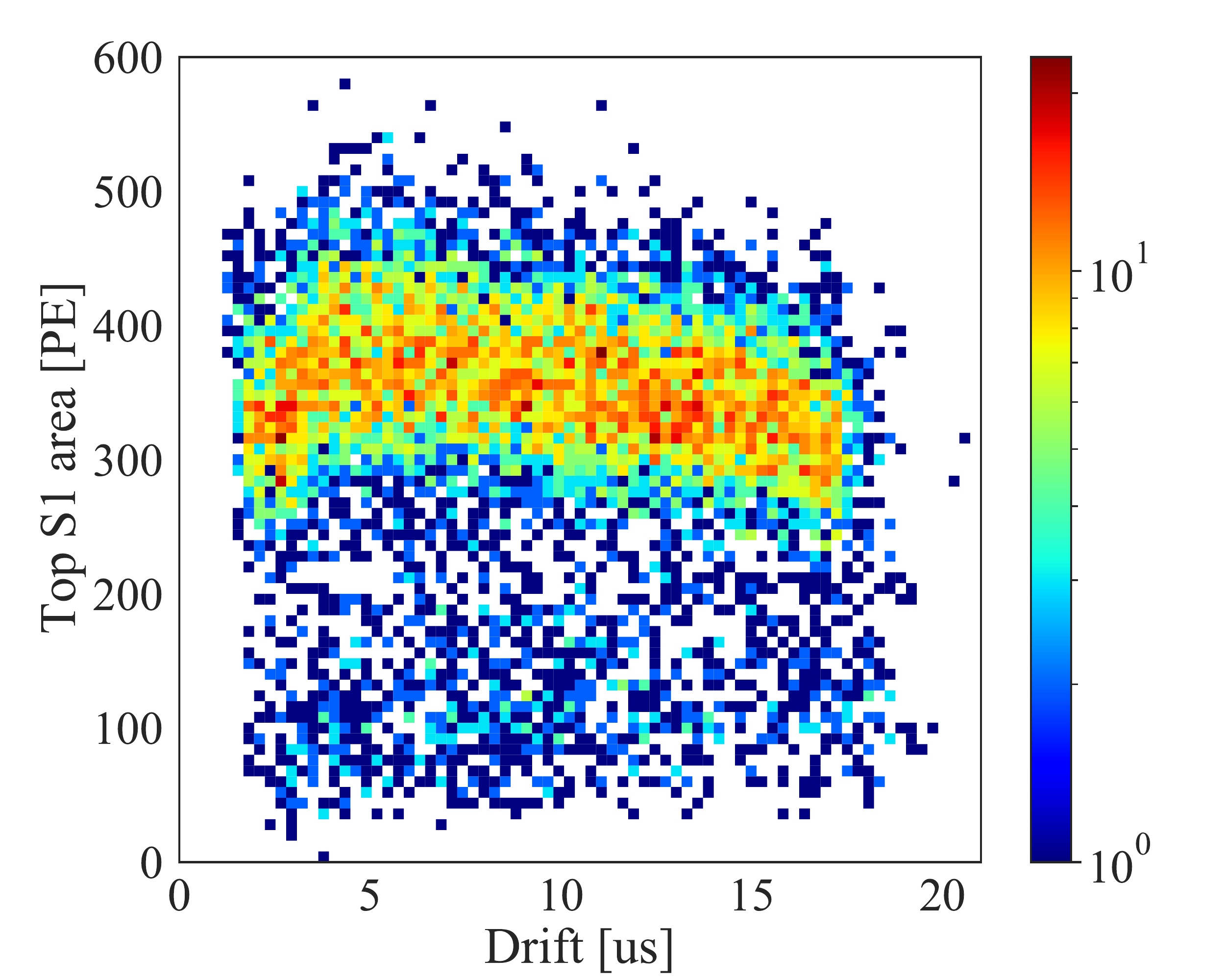}
\end{minipage}
\begin{minipage}{.45\textwidth} 
  \centering
  \includegraphics[width=1.04\linewidth]{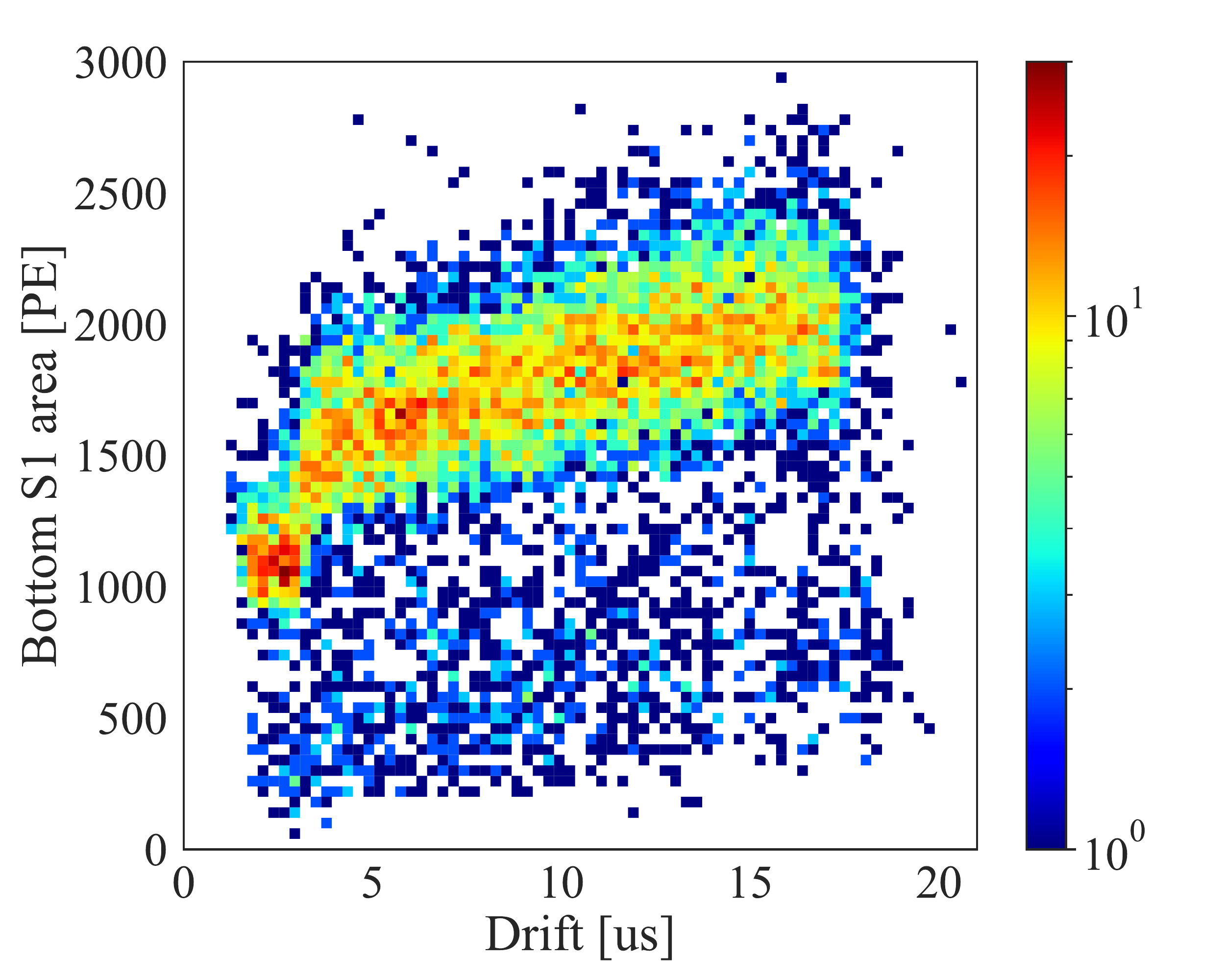}
\end{minipage}
\caption{S1 area vs. event drift time in the top (top plot) and bottom (bottom plot) PMT arrays for full energy deposition gammas from a $^{137}$Cs calibration run.  As the drift time increases, corresponding to events closer to the bottom of the active volume, the top (bottom) PMT array sees less (more) light. The reduced S1 area at drift times $<3.5$ $\mu$s are events in the extraction region, where a lower scintillation yield is expected due to the stronger electric field.}
\label{fig:s1z}
\end{figure}

To measure the detector's efficiency for collecting scintillation light, several external gamma sources, including $^{133}$Ba (356 keV), $^{22}$Na (511 keV), and $^{137}$Cs (662 keV), were used.
The gamma energies above were exclusively used for this calibration because 
lower-energy gamma rays are not able to penetrate into the active TPC volume, while MeV-scale gammas would produce scintillation pulses that can saturate the PMTs and electronics. 
To avoid saturation of the electroluminescence-amplified S2 signals, the TPC electron amplification field was set to be just above the onset of electro-luminescence in the gas phase and held constant for all calibration data runs.  While the TPC electron gain in these calibrations is different from that used for the neutron scatter measurement, the scintillation light collection efficiency (LCE) is an intrinsic property of the detector and  
can be measured using the gamma source data.

Fig.~\ref{fig:s1z} shows the S1 pulse areas observed by the top and bottom PMTs for $^{137}$Cs gammas with full energy depositions in the xenon TPC.
Because photons emitted in the liquid can undergo total internal reflection at the liquid-gas boundary, most S1 light is collected by the bottom PMT and only $\sim$16\% is detected by the top PMT array. 
A strong S1 dependence on the depth of interaction is also observed ($\sim$20\% across the full drift region), with more (fewer) photoelectrons collected by the bottom PMT (top PMT array) for events near the cathode.  Prior to estimating the LCE with calibration data,  we used empirical polynomial functions to separately correct the top and bottom PMT S1 signal amplitudes to remove the S1 area position dependence.  The reference point for this correction was chosen to be approximately the middle of the active volume.
The same correction is also applied to the xenon recoil data. 

To estimate the LCE after position corrections, we produced a Doke plot \cite{DobiThesis} by performing two-dimensional Gaussian fits to the peak-of-interest in each gamma source dataset.   
The LCE (denoted as $g1$ below) can then be 
calculated using the following relationship between energy and observables:
\begin{equation}
    E = W\left(n_{ph} + n_{e}\right) = W_{e/\gamma}\left(\frac{S1}{g1} + \frac{S2}{g2}\right)
\end{equation}
where $W_{e/\gamma}$ is the work function (fixed at 13.7 eV/quanta for this analysis as measured in \cite{Doke:2002oab,DahlThesis}), $n_{ph}$ is the number of scintillation photons produced, $g2$ is the electron gain parameter, and $n_{e}$ is the number of ionization electrons produced.

\begin{figure}[!ht]
    \centering
  \includegraphics[width=0.94\linewidth]{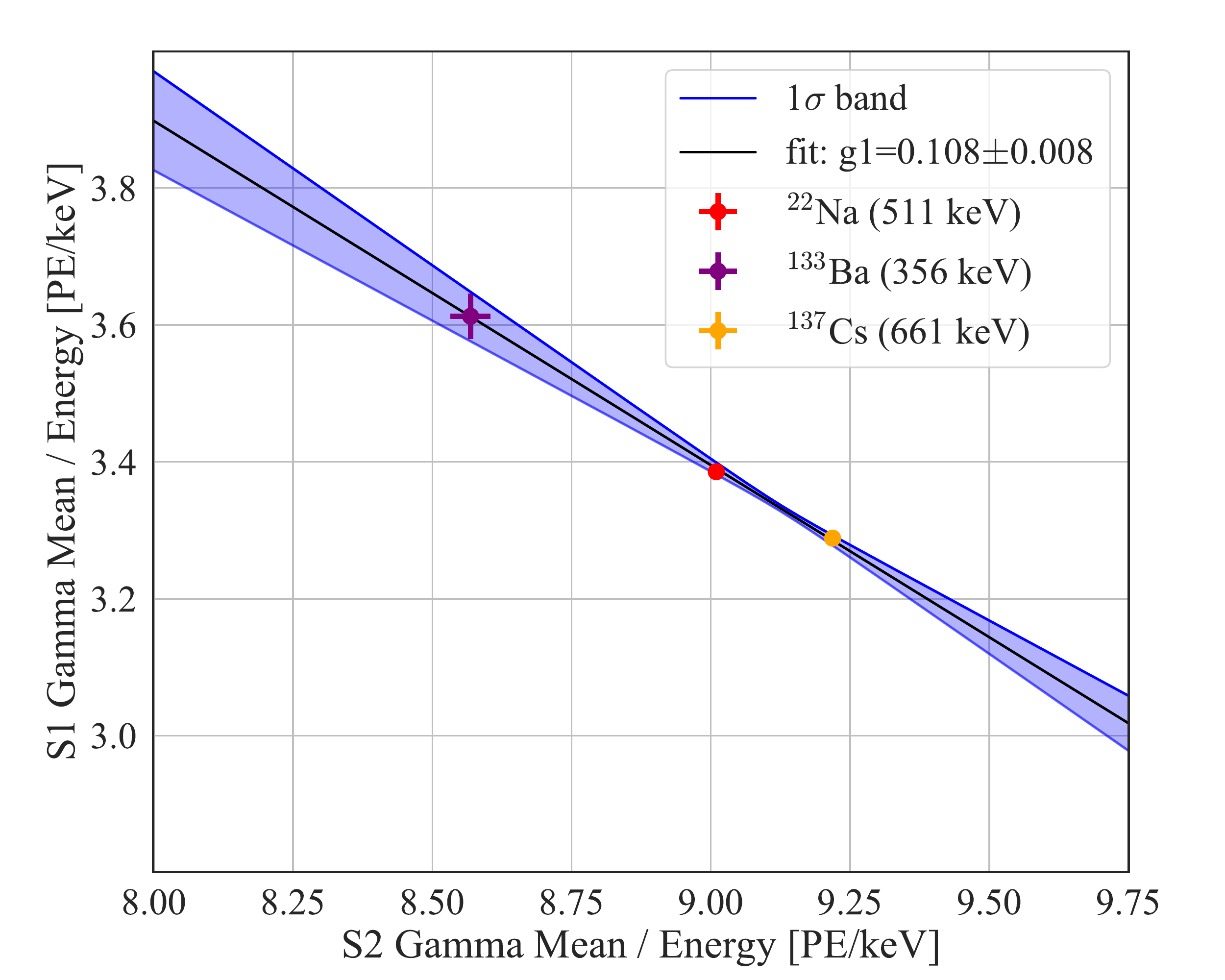}
\caption{Doke plot composed of data points from gamma calibrations.  The blue shaded region shows the 1$\sigma$ uncertainty band on the line fit.}
    \label{fig:LCE}
\end{figure}

As illustrated in Fig.~\ref{fig:LCE}, the above expression can be rearranged to yield an anticorrelated line between $S1/E$ and $S2/E$ for different calibration sources with $g1$ as a parameter that can be determined via a fit.  The LCE was measured to be $0.108\pm0.008$ PE/photon for the xenon TPC.  
The $^{137}$Cs and $^{22}$Na data points carry relatively small uncertainties which are determined by the fit to the endpoint.  The $^{133}$Ba data point has larger uncertainties due to two additional systematic effects.  First, the main $^{133}$Ba peak at 356 keV (62.1\% intensity) overlaps with the Compton shoulder.  This uncertainty is quantified by repeating the 2-D Gaussian fit while varying the lower-energy fit boundary. The second uncertainty comes from possible contamination of a neighboring higher-energy peak at 383 keV (8.9\% intensity). 
The level of bias was estimated to be at the $<$1\% level with a NEST simulation assuming the detector's conditions during calibration.

\subsection{Simulation of scintillation/ionization spectra}\label{subsec:Simulation}

The TPC xenon recoil energy distribution in coincidence with each LS detector is simulated using Geant4.  
The DT source, lead and borated water shielding, xenon TPC, and backing detectors are all modeled using the BACCARAT simulation package~\cite{AKERIB201263}.
All primary DT Neutrons are generated with 14.1 MeV kinetic energy at the DT source interaction point, and initial directions are drawn isotropically within a cone of 1 degree angle directed toward the collimation hole. 
Energy deposits in the backing detectors and active xenon volume from the primary neutron and subsequent particles produced in each simulated event are recorded and stored for analysis.  Cuts equivalent to those described in Sec.~\ref{subsec:EventSelection} are then applied to the simulated data. 
Because LS pulse shape is not implemented in the simulation, we require energy depositions in the LS detectors to originate from a neutron scatter.

The energy spectra from the simulation passing the cuts are then used to 
generate S1 and S2 responses that can be compared with the data. 
The number of photons produced by a certain xenon recoil energy is sampled from a Gaussian distribution with mean
\begin{equation}
    \mu_{ph} = E \times L_{y}(E)
\end{equation}
and standard deviation
\begin{equation}
    \sigma_{ph} = W \sqrt{E \times L_{y}(E)}
\end{equation}
where $E$ is the xenon recoil energy, $L_y$ is the energy-dependent light yield (number of photons produced per unit of recoil energy deposited in the xenon), and $W$ is an empirical width parameter which models interaction dynamics in xenon (such as electron/ion recombination fluctuations and Fano statistics \cite{LUXTritiumCalibration_2016,DOKEFano1976353}).  

The light yield $L_y$ in each dataset's peak region is modeled using two different methods. The first method simply treats the light yield as a constant for each coincidence data set. This is a good approximation for most xenon recoil energies when the peaks are relatively narrow, but overestimates the S1 spread for other coincidence angles with broader energy distributions. To account for the energy dependence of the xenon recoil yield for these broad peaks, we implement a second method that models $L_y$ in the form of $L_y(E)=B\times E^{\tau}$ in the vicinity of the peak energy, where $B$ and $\tau$ are free parameters in the model. 
The power-law form allows the xenon recoil yield to vary modestly with energy without losing generality, and this assumption is qualitatively consistent with our measurement results presented in Sec.~\ref{sec:discussion}.  Although both methods agree within the model fit uncertainties, the light and charge yield results from the power-law form are presented below as they provide better fits to the data.

The detected number of photoelectrons is then modelled with a binomial distribution with the LCE probability measured in Sec. \ref{subsec:EEELCE}.  The photoelectron count is further smeared with the single photoelectron resolution measured in data, with a relative width of $\sigma_{SPE}=58\%$.  Finally, the sampled photoelectron values are binned and scaled with an overall normalization factor $A$ before being compared to data. 

The development of the S2 distribution model follows the same procedure as above and also uses a power-law form, only the light yield $L_y(E)$ is replaced with $Q_y(E)$, the LCE is replaced with EEE, and $\sigma_{SPE}$ is replaced with the measured single electron resolution $\sigma_{SE}=26.7\%$.

\begin{figure*}[!ht]
\centering
\includegraphics[width=1.00\linewidth]{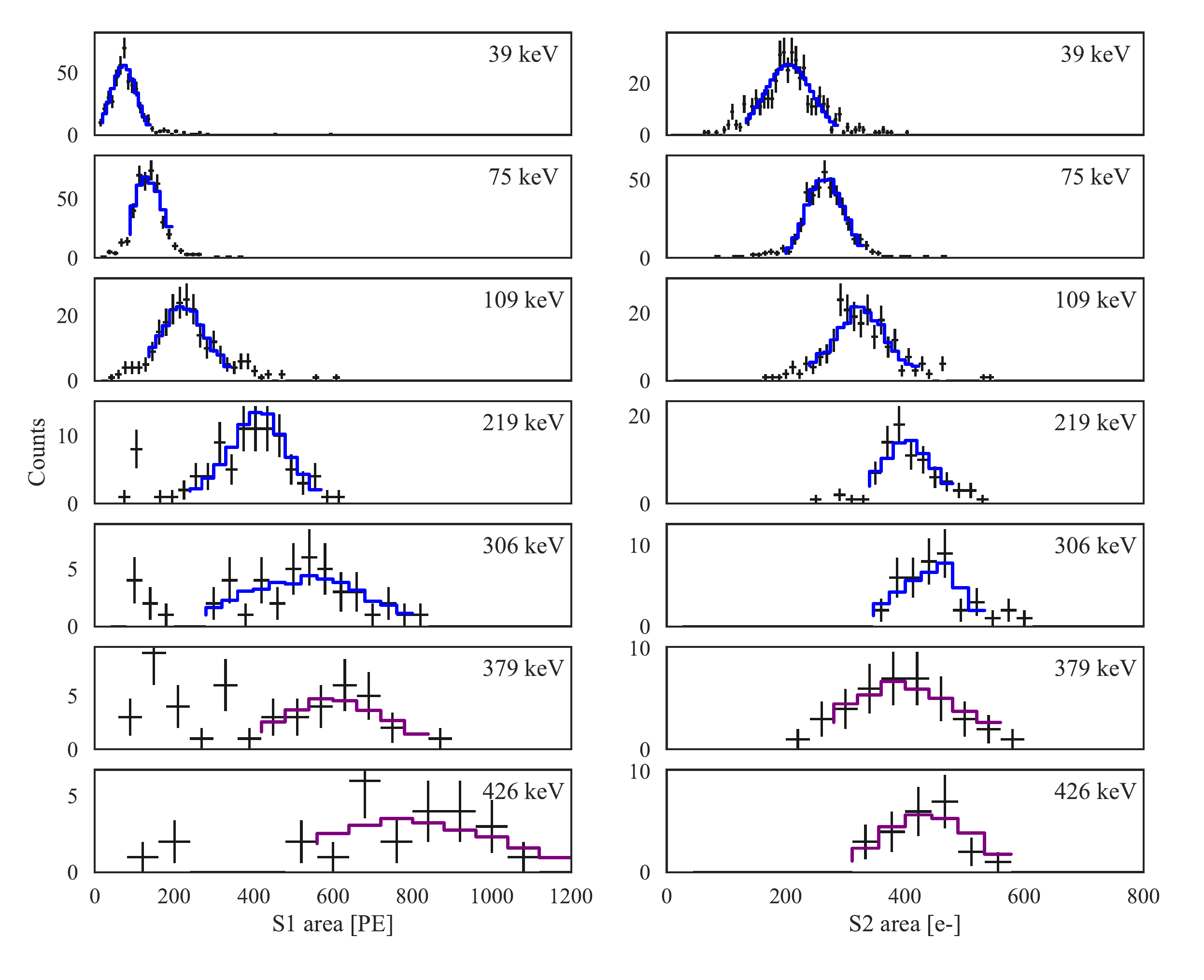}
\caption{Measured S1 and S2 spectra for xenon recoil candidate events at different energies (black histogram), along with best model fits (blue curve for 2000 V/cm drift field data, purple curve for field-averaged data). There is a downward S2 peak shift for the field-averaged data due to the effective field being lower than 2000 V/cm. 
All recoil energy fits are performed in the $[-2\sigma,+2\sigma]$ range around the best fit Gaussian of the peak in data.}
\label{fig:peakfits}
\end{figure*}

\subsection{Fit methodology}\label{subsec:FitMethodology}

The S1 and S2 distribution models are compared to the measured data to extract the $L_y$ and $Q_y$ yields as a function of xenon recoil energy. This process uses a Markov Chain Monte Carlo (MCMC)-based approach to sample the $L_y$ and $Q_y$ parameter space, determine the best fit parameters, and estimate their uncertainties.  We use a likelihood function to quantify how well a given model agrees with the measured data.  Assuming each bin in the data distribution is Poisson-distributed,  the log-likelihood function for a scintillation spectrum is
\begin{equation}
    \mathcal{L}\left(n|m(L_{y}\left(B,\tau\right),W,A)\right) = \sum_{j=1}^{N} m_j \ln{n_j} - n_j - \ln{(m_j !)} 
\end{equation}
where $n_j$ and $m_j$ are the $j$th values in the binned data and model distributions, respectively.  For the ionization spectra analysis,  $L_y$ is replaced with $Q_y$.  

The best-fit model is determined by sampling the model parameter space using the Metropolis-Hastings algorithm.  
As the algorithm steps around the model's parameter space, the most likely S1 or S2 model fit parameters can be found by maximizing the likelihood function.  Once near the maximum likelihood, the Monte Carlo component of each step will occasionally allow the algorithm to step away from the maximum and explore the neighboring parameter space, which is used to estimate the uncertainty on the best fit model.  The best fit value for each parameter is calculated as the median of the scanned values after the algorithm has converged, and the uncertainty is reported as the 68\% quantile spread around the median. Each analysis was ran with 10000 steps, and the mean/uncertainty calculations ignore the first 1000 steps while the fit converges.

\section{Results and discussions}
\label{sec:discussion}

\begin{table*}[ht]
\centering
\caption{Summary of light and charge yields measured in this work as a function of neutron scattering angle/xenon recoil energy.  The reported recoil energy is the estimated recoil energy mean from simulation using a Gaussian fit.  The reported light/charge yield are the values calculated using the model's best-fit $L_y(E)$ and $Q_y(E)$ functions at the energy peak. The $L_y$ uncertainties shown in the table are the MCMC fit uncertainties.  The $Q_y$ uncertainties shown in the table are the total uncertainty from the MCMC fits and the variation in S2 size seen for different extraction voltages (explained in text).  The field-averaged column is the best fit value when analyzing the sum of all data at all drift and extraction voltage settings.}

\vspace{0.2cm}

\footnotesize
\begin{tabular}{|C{0.085\linewidth}|C{0.09\linewidth}|C{0.072\linewidth}|C{0.072\linewidth}|C{0.072\linewidth}|C{0.072\linewidth}|C{0.072\linewidth}|C{0.072\linewidth}|C{0.072\linewidth}|C{0.072\linewidth}|C{0.072\linewidth}|C{0.072\linewidth}|}
\hline
scattering & recoil & \multicolumn{5}{|c|}{Qy} & \multicolumn{5}{|c|}{Ly} \\
\cline{3-12}
angle (deg.)  & energy (keV)  & 0.2 kV/cm & 0.76 kV/cm & 2.0 kV/cm & Field avg. & TOF sys. & 0.2 kV/cm & 0.76 kV/cm & 2.0 kV/cm & Field avg. & TOF sys.\\
\hline
\hline

$36 \pm 1$ & $39 \pm 3$ & $4.39^{+0.23}_{-0.33}$ & $4.63^{+0.52}_{-0.50}$ & $5.04^{+0.50}_{-0.48}$ & - & $^{+5.3\%}_{-2.7\%}$ & $16.2^{+0.8}_{-0.7}$ & $15.2^{+0.5}_{-0.8}$ & $15.7^{+0.8}_{-0.7}$ & - & $^{+5.5\%}_{-2.3\%}$\\
\hline
$50\pm1$  & $75 \pm 4$ & $3.26^{+0.16}_{-0.16}$ & $3.54^{+0.17}_{-0.17}$ & $3.69^{+0.14}_{-0.14}$ & - & $^{+1.6\%}_{-1.4\%}$ & $16.8^{+0.2}_{-0.2}$ & $16.4^{+0.2}_{-0.2}$ & $16.4^{+0.3}_{-0.3}$ & -& $^{+1.9\%}_{-1.3\%}$  \\
\hline
$67\pm2$ & $109 \pm 6$ & $2.62^{+0.16}_{-0.16}$ & $2.87^{+0.15}_{-0.15}$ & $3.10^{+0.15}_{-0.15}$ & - & $^{+4.3\%}_{-2.3\%}$ & $18.3^{+0.5}_{-0.4}$ & $18.3^{+0.5}_{-0.5}$ & $18.8^{+0.5}_{-0.4}$ & - & $^{+3.5\%}_{-0.5\%}$\\
\hline
$92\pm2$ & $219 \pm 7$ & $1.66^{+0.08}_{-0.08}$ & $1.82^{+0.09}_{-0.10}$ & $1.93^{+0.08}_{-0.08}$ & - & $^{+6.4\%}_{-1.8\%}$ & $17.9^{+0.4}_{-0.4}$ & $17.7^{+0.3}_{-0.4}$ & $17.6^{+0.5}_{-0.4}$ & - & $^{+2.2\%}_{-0.1\%}$\\
\hline
$115\pm3$ & $306 \pm 10$ & $1.25^{+0.08}_{-0.08}$ & $1.45^{+0.13}_{-0.12}$ & $1.59^{+0.07}_{-0.07}$ & - & $^{+2.0\%}_{-1.2\%}$ & $18.2^{+0.6}_{-0.5}$ & $16.1^{+0.7}_{-0.6}$ & $18.0^{+1.0}_{-0.8}$ & - & $^{+6.4\%}_{-3.0\%}$ \\
\hline
$140\pm2$  & $379 \pm 5$ & - & - & - & $1.12^{+0.14}_{-0.14}$ & $^{+3.1\%}_{-2.0\%}$ &- & - & - & $15.7^{+0.9}_{-0.8}$ &$^{+8.0\%}_{-3.1\%}$\\
\hline
$162\pm2$  & $426 \pm 2$ & - & - & - & $1.11^{+0.14}_{-0.13}$ &$^{+4.6\%}_{-1.2\%}$ & - & - & - & $18.5^{+1.3}_{-1.5}$ &$^{+4.9\%}_{-10.0\%}$\\

\hline
\hline
\multicolumn{2}{|c|}{LCE systematic unc.} & \multicolumn{5}{c|}{-} & \multicolumn{5}{c|}{$\pm7.4\%$} \\
\hline
\multicolumn{2}{|c|}{EEE systematic unc.}  & \multicolumn{5}{c|}{$\pm3.0\%$} & \multicolumn{5}{c|}{-}\\
\hline
\end{tabular}
\label{tab:Results}
\end{table*}

The measured S1 and S2 spectra for xenon recoil energies up to 306 keV for the drift field of 2000 V/cm (best fit models in blue curves) are shown in Fig.~\ref{fig:peakfits}. 
At high xenon recoil energies it was observed that the elastic neutron scatter rates are significantly lower than the prediction of Geant4 simulations or direct calculations using evaluated nuclear data~\cite{NeutronXCs}. This rate deficiency is also found to be present for some medium xenon recoil energies with well-defined neutron TOF structures and does not appear to originate from an efficiency loss associated with the analysis cuts. Due to low statistics at 379 and 426 keV recoil energies, data taken at all three drift settings for these recoil energies were combined to produce a single set of spectra, and the light and charge yields are reported as field-averaged values. The best fit models for these two energy points are shown in purple in Fig.~\ref{fig:peakfits}.

The fits are carried out only in the vicinity of the main S1 and S2 peaks, from 2 standard deviations below the peak mean to 2 standard deviations above.  
Prior to performing the S1 (S2) fit, all events with an S2 (S1) size 3 standard deviations below the mean are removed to mitigate low energy background contamination as illustrated in Fig.~\ref{fig:s12}. The same cut is applied to data shown in Fig.~\ref{fig:peakfits} for consistency. 

For the 39 keV recoil energy, a good model fit cannot be obtained due to the prediction of a larger-than-observed low-energy background in the Geant4 simulation.    
This background is due to neutrons that undergo scattering in passive materials around the TPC in addition to an interaction inside the TPC. However, this nuclear recoil energy coincides with the $\sim$39 keV resonance of $^{129}$Xe and is predicted to have a reduced elastic neutron scatter cross section. 
Given the lack of direct nuclear interaction data and the large anticipated uncertainty of the evaluated interaction cross sections for 14.1 MeV neutron scattering off xenon, we chose to remove the multiple-scatter background in the simulation, which is equivalent to increasing the elastic scatter cross section at this energy, while carrying out the MCMC fit of the signal model to the data.  
To evaluate the uncertainty resulting from this approach, we model the 
S1 and S2 distributions at this energy as single Gaussians and calculate the light and charge yields directly by dividing the mean observed quanta (corrected with LCE and EEE) by the kinematically calculated recoil energy.    
The difference in the yield values between this approach and the MCMC fit is treated as a systematic uncertainty that is added to the fit uncertainty returned by the MCMC. The Gaussian fit is presented as the best fit in Fig.~\ref{fig:peakfits}.

\subsection{Experimental uncertainties}\label{subsec:systematics}

Statistical uncertainties for the measurement results are captured by the MCMC fit procedure as explained in Sec.~\ref{subsec:FitMethodology}. 
Additional yield uncertainties at each recoil energy can result from uncertainties in the placement of the corresponding backing detector with respect to the xenon TPC.  
The uncertainties of the measured backing detector positions are used to evaluate the uncertainty in the scattering angle, which is then propagated to an uncertainty in recoil energy based on neutron elastic scattering kinematics, as summarized in Table \ref{tab:Results}.

\begin{figure*}[!ht]
\centering
\begin{minipage}{.49\textwidth}
  \centering
  \includegraphics[width=1.0275\linewidth]{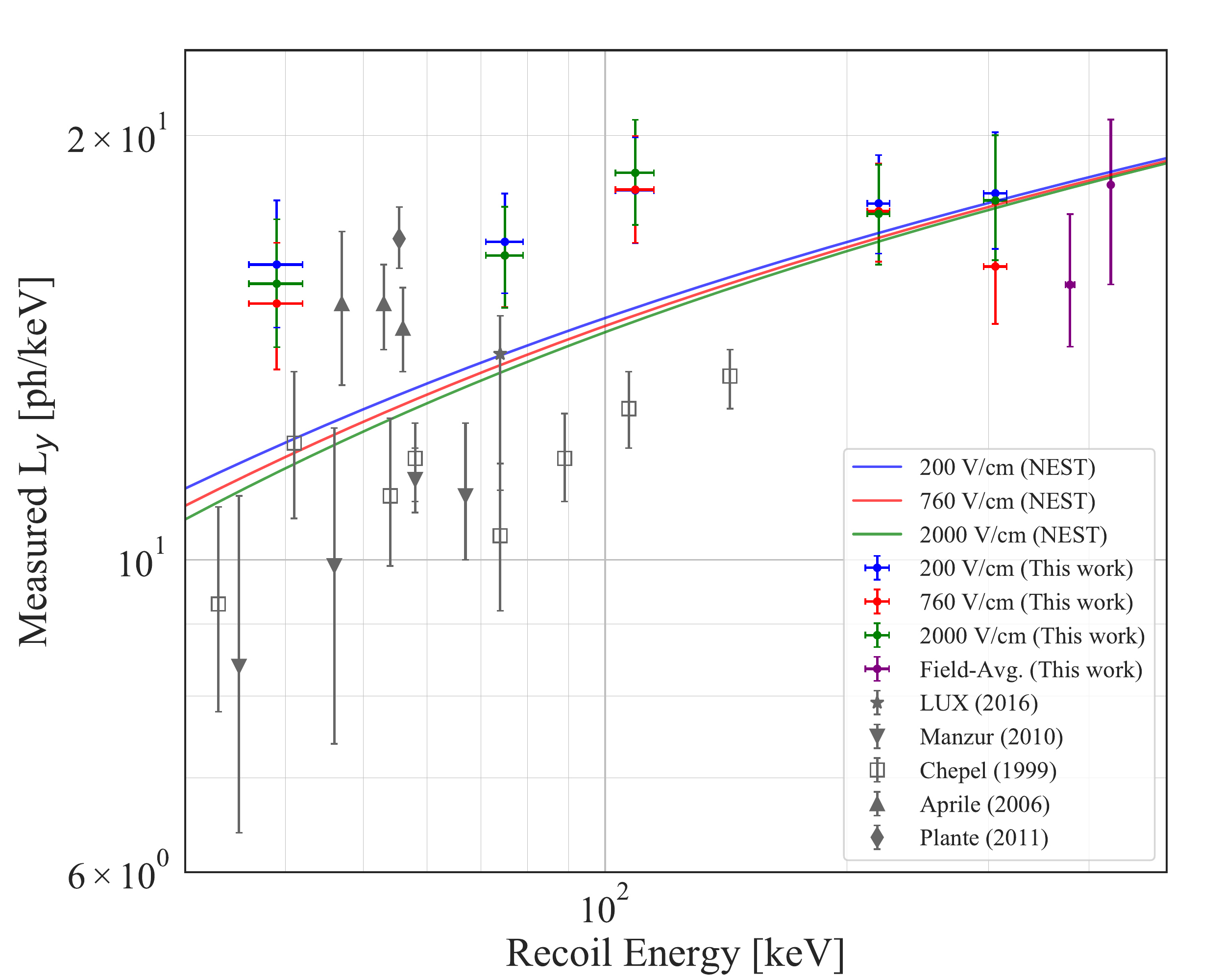}
\end{minipage}
\begin{minipage}{.50\textwidth}
  \centering
  \includegraphics[width=1.00\linewidth]{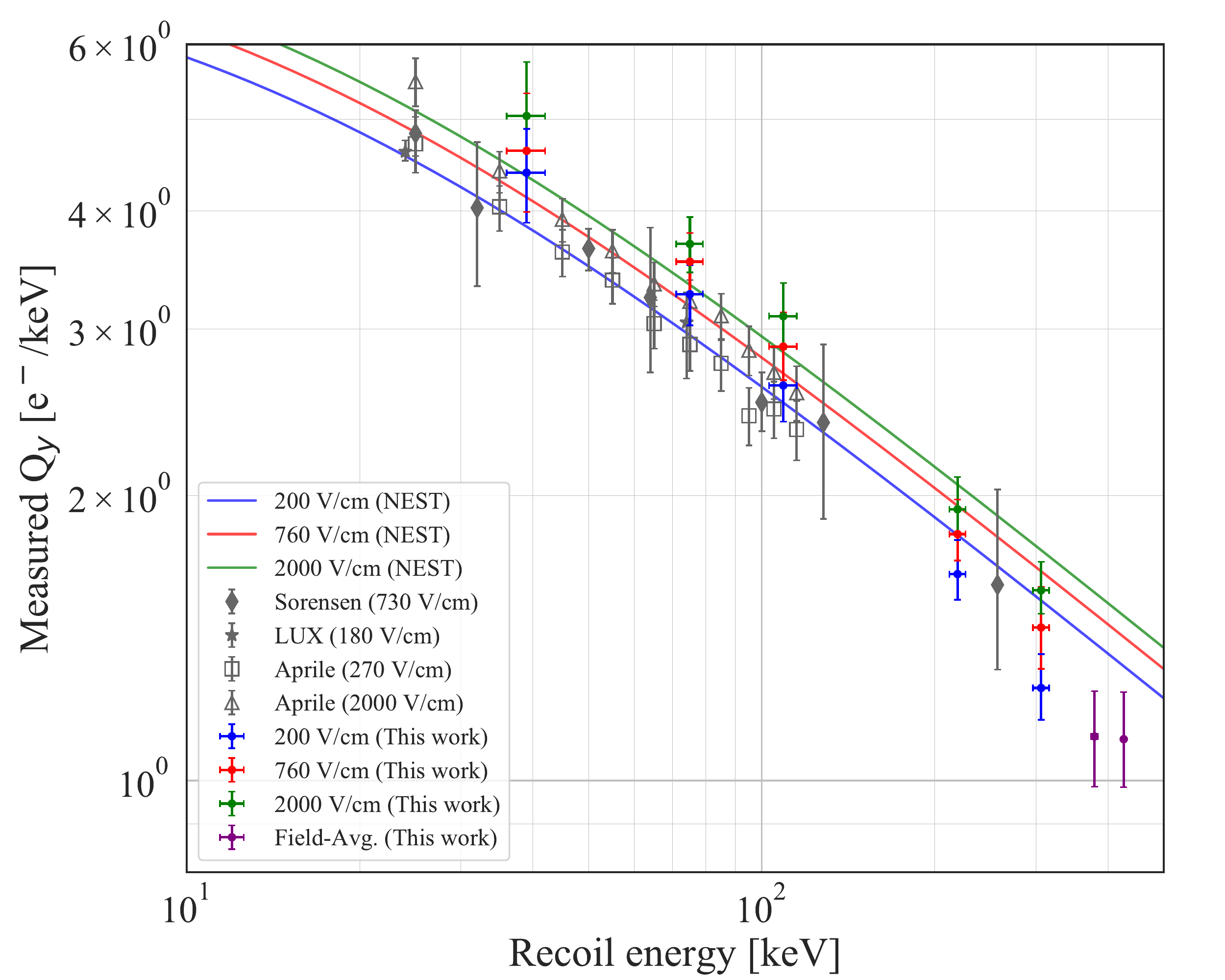}
\end{minipage}%
\caption{Light and charge yields for xenon recoils and their uncertainties measured as a function of energy in this work.  Error bars reflect the combined statistical and systematic uncertainties for each measured data point.  The current predictions from the NEST model are overlaid for comparison, as well as measurements from \cite{Manzur2010,Chepel1999,Aprile2006,Plante2011,AprileIon2006,SORENSENHighE,LUX:2016ezw,SORENSEN2009339} as compiled in \cite{Lenardo2015_XeLightCharge}.}
\label{fig:lqresults}
\end{figure*}

For the light yield results, the uncertainty on the TPC LCE, described in Sec.~\ref{subsec:EEELCE}, applies equally to the results at all recoil energies.
The charge yield estimates have two main contributors to the overall uncertainty: the electron extraction efficiency (EEE), and a discrepancy in S2 size distributions measured at the same energy but under different extraction field settings. 
For the measurements at drift fields of 200 V/cm and 760 V/cm, data was taken both before and after the change of operation conditions amid the experiment, as described in Sec.~\ref{sec:setup}. 
During this process, the TPC drift fields were maintained at the same levels and the yield values are not expected to change. Due to unknown sources, however, a discrepancy of $\sim$4\% in the charge measurements was observed between the data before and after the voltage change even with all efficiencies separately evaluated and corrected for. As a result, the charge yields are reported as the average values from both datasets, with the difference treated as a systematic uncertainty.  This additional uncertainty is not applied to the light yield measurements, as the S1 size is independent of the extraction field and was not observed to shift after the voltage changes.  The final results of the ionization and scintillation yields are summarized in Table \ref{tab:Results}.  

As explained in Sec.~\ref{subsec:Simulation}, the simulations only consider a small fraction of the neutrons emitted in a narrow cone by the DT source, and neutrons emitted at large angles may scatter in the shielding materials and produce a background. 
In addition, neutrons may scatter inelastically with xenon or undergo (n, 2n) processes and still produce coincidence events with lower energy neutrons traveling to the backing detectors. 
To quantify these effects, the analysis was repeated using both a TOF cut window [-2,0] ns around the estimated TOF mean, which should have minimal low energy neutron background contamination, and a TOF cut window [0,+2] ns around the TOF mean, which could have a larger presence of low energy neutron background contamination.  The largest differences in the estimated light and charge yield measured at different drift fields and in the field-averaged data are propagated into the final result as a systematic uncertainty.

\subsection{Comparison of results to literature}

The measured light and charge yield values as a function of xenon recoil energy in this experiment are summarized in Fig.~\ref{fig:lqresults}.  The yield values predicted by NEST v2.3.6~\cite{Szydagis_2011} and those reported by previous measurements~\cite{HORN2011471,LUX:2016ezw,XENON100:2013smi,Lenardo2015_XeLightCharge,Manzur2010,Chepel1999,Aprile2006,Plante2011,AprileIon2006,SORENSENHighE,SORENSEN2009339} are also shown for comparison.

The light yield values are observed to have a relatively mild dependence on xenon recoil energy in the energy region studied.  
Our results are slightly higher than NEST predictions below 109 keV, and agree with some past measurements. Due to the measurement uncertainties, we do not observe a significant dependence of the light yield on the drift electric field, which is also consistent with the predictions in NEST.

The charge yield measured in this work decreases as a function of recoil energy across the entire 39--426 keV energy range.  Values measured in this work are in good agreement with other measurements in overlapping energy windows. We comment that the continually decreasing trend in the charge yield can have an implication for energy reconstruction of xenon nuclear recoil events.  Since the charge yield approximately falls inversely proportional to recoil energy, the S2 size is almost stationary for most energies above 200 keV, as can be seen in Fig.~\ref{fig:peakfits}; as a result, energy reconstruction for higher energy recoils would rely primarily on the S1 size.   

A modest increase of the charge yields at stronger drift fields is observed.  This trend is expected, as increasing the drift field strength increases the probability that ionized electrons will migrate away from the ionized cloud and escape recombination.  The amount of increased charge yield as a function of drift field is also in good agreement with NEST predictions.

\section{Conclusion}
\label{sec:conclusion}

We deployed a 14.1 MeV neutron beam to characterize the light and charge yields of xenon nuclear recoils in a dual-phase xenon TPC.  Liquid scintillator coincidence detectors were placed around the xenon TPC at angles between 36-162 degrees to tag neutron elastic scatters between energies of 39-426 keV in the TPC's active xenon volume.  By selecting neutron elastic scattering candidates using TPC and backing detector-based event selection cuts and fitting the data with a simulation-based detector signal model, we measured the light and charge yield in liquid xenon as a function of drift field for xenon recoil energies between 39 and 306 keV.  By summing all drift field data, field-averaged light and charge yields were also measured for the 379 and 426 keV recoil energies.  This result will improve the accuracy of nuclear recoil modeling in liquid xenon and enable new dark matter searches to be carried out in the hundreds of keV energy region in xenon TPC experiments.

\begin{acknowledgments}
This project is supported by the U.S. Department of Energy (DOE) Office of Science, 
 Office of High Energy Physics under Work Proposal Number SCW1676 and SCW1508 awarded to Lawrence Livermore National Laboratory (LLNL). 
 LLNL is operated by Lawrence Livermore National Security, LLC, for the DOE, National Nuclear Security Administration (NNSA) under Contract DE-AC52-07NA27344.
 D.~Naim, J.~Kingston, V. Mozin and P. Kerr are partially supported by the DOE/NNSA under Award Number DE-NA0000979 through the Nuclear Science and Security Consortium. 
B. G. Lenardo is supported by the DOE Office of nuclear physics.
 We thank Tony Sorensen and Sean Mok for a series of constructive discussions on the design and construction of the neutron shielding structure, which is critical for this experiment. 
 
 LLNL IM release number: LLNL-JRNL-837368
 
\end{acknowledgments}
\typeout{}
\bibliography{biblio}
\bibliographystyle{apsrev}

\end{document}